\documentclass[twocolumn,english,aps,pre,floatfix]{revtex4}
\usepackage[T1]{fontenc}
\usepackage[latin9]{inputenc}
\usepackage{textcomp}
\usepackage{amstext}
\usepackage[dvips]{graphicx}



\usepackage{babel}
\begin{document}

\title{Dielectric permittivity of deeply supercooled water according 
to the measurement data at the frequencies 7.6~GHz and 9.7~GHz}

\author{G. S. Bordonskiy, A. A. Gurulev, A. O. Orlov}

\affiliation{Institute of Natural Resources, Ecology and Cryology SB RAS, 16 Nedorezova,
p/b 1032, 672002 Chita, Russia}

\email{lgc255@mail.ru}

\begin{abstract}
Dielectric permittivity of supercooled volume water has been measured 
in the range of temperatures from $-20\,^{\circ}$C to $-60\,^{\circ}$C 
at the frequencies $7.6$~GHz and $9.7$~GHz. 
The measurements have been made using microwave resonators and wetted 
silicate sorbents. 
From the data obtained, the temperature dependences of the relaxation 
frequencies were updated for the Debye model with two relaxation 
frequencies. 
The updated formulae for dielectric permittivity of cold water may be 
applied to the frequency range $7\ldots200$~GHz.
\end{abstract}

\maketitle

\section*{Introduction}

The knowledge of dielectric permittivity of supercooled volume water 
(down to $-60\,^{\circ}$C) at microwaves is of interest for the study 
of the physical and chemical properties of water and of water solutions. 
There are numerous problems relating to the study of hydrosphere, 
biosphere, and atmosphere, where the knowledge of the microwave 
characteristics of water and their details is required~\cite{MWen2004, Ros2015, TKC2016}. 

However, measuring the complex relative dielectric permittivity 
($\dot{\varepsilon}$), especially at the temperatures ($T$) below 
$-20\,^{\circ}$C, is a complicated task~\cite{Ros2015}. 
In~\cite{BCS1982}, it was solved using microemulsion of water, when 
the researchers were able to cool the water samples down to 
$-18\,^{\circ}$C and to make measurements of the real 
($\varepsilon^{\prime}$) and imaginary ($\varepsilon^{\prime\prime}$) 
parts of $\dot{\varepsilon}$ near the frequency $9.6$~GHz. 

In~\cite{BOK2019, BO2019a} wetted nanoporous silicate materials were 
used for the purpose, in which water could be supercooled to 
$-90\,^{\circ}$C~\cite{LCh2012}. 
Dependences $\varepsilon^{\prime\prime}$ were found in a broad range of 
frequencies ($f$) and temperatures. 
In those works, originally dependence of the attenuation constant for 
radiation intensity of wetted medium was measured. 
Further, to calculate $\varepsilon^{\prime\prime}$, theoretical values 
$\varepsilon^{\prime}$ were used, which, however, were not experimentally 
validated for the water temperature below $-20\,^{\circ}$C. 
The situation was complicated also by the fact that, according to the 
study results~\cite{BO2019b} of the dielectric permittivity models~\cite{MWen2004, Ros2015, Ell2007}, 
which were applied to the temperature range below $-20\,^{\circ}$C, 
it was found that the values of $\varepsilon^{\prime}$ differed by several 
times.

To objective of the present study was to measure and to investigate 
the temperature dependence $\varepsilon^{\prime}$ for pore water, close 
for its characteristics to supercooled metastable volume water. 
The study was conducted for the temperature range $0\ldots-60\,^{\circ}$C 
near the frequencies $7.6$ and $9.7$~GHz.

\section*{The method used}

As shown in~\cite{MMF2017, CGAD2011, SLSM2011}, water contained in 
nanometer-size pores of silicate sorbents has properties differing 
from the volume water only for $1-2$ layers of water molecules on 
the pore surface. 
The other layers are close for their characteristics to metastable 
volume water. 
This unique feature of the silicate sorbents was used for measuring 
$\varepsilon^{\prime}$ of supercooled water. 

It is known that in such media the phase transition temperature ($T_c$) 
decreases by the value ${\Delta}T_c=c/(R-t)$, where $c=62$~degrees$\cdot$nm, 
$R$~-- the pore radius in nm, $t\approx0.38$~nm~\cite{SLSM2011}. 
In the case of pores having the radius $3$~nm, ${\Delta}T_c$ is 
$\approx24\,^{\circ}$C. 
In the case of incomplete filling of the pores, additional supercooling 
is observed, compared to ${\Delta}T_c$ by approximately 
$10\ldots15\,^{\circ}$C~\cite{GAAA2016}. 
In addition, in the cooling --- heating process, the freezing --- 
melting temperature hysteresis is formed, at which the freezing 
temperature is lower than the values found from the formula for 
${\Delta}T_c$. 
The hysteresis may exceed $10\,^{\circ}$C and is determined by the 
value of the average water content and the cooling rate~\cite{GAAA2016, W2010}.

To reach the temperature in the area of the ``no man's land'' 
($T=-37\ldots-120\,^{\circ}$C)~\cite{LCh2012, GAAA2016} low values of 
wetted sorbent materials with the gravity water content $\sim3\ldots4$\% 
were used. 
The cooling rate was selected not to exceed $10\,^{\circ}$C/hr. 

As the proportion of volume water in the medium is $\sim1$\%, and its 
value $\varepsilon^{\prime}$ goes down with the temperature decrease, 
the resonator method was used, in which complete filling of the 
resonator with the medium is achieved. 
Such a method was used, for example, in~\cite{MatWeg1987} to measure 
$\varepsilon^{\prime}$ of freshwater ice in determining the impact of 
small concentrations of liquid inclusions in it. 
In the course of our study, we discovered the ability of wetted 
disperse media to form macroscopic inhomogeneities due to migration 
of the liquid in the presence of a temperature gradient and at freezing 
of free water between the grains. 
This effect leads to distortion of the resonance curves and to 
essential reduction of the measurement accuracy. 
In~\cite{BOK2019}, it was eliminated at measurement of $\varepsilon^{\prime\prime}$, 
placing large--size samples in free space, which resulted in reduction 
of the impact of the medium inhomogeneities and of the spatial 
dispersion caused by them. 

A similar approach was used in this study for resonator measurements. 
Resonators were used with higher numbers of the oscillation modes and 
hence of the resonator volume. 
Such a technique allows averaging by a large number of inhomogeneities. 
In addition, data averaging was performed at the frequency range $\sim1$\% 
of the average frequency.

In our study, we used rectangular transmission-type resonators, connected 
to waveguide transmission lines with the waves of the $H_{10}$ type. 
The resonance frequencies of a rectangular resonator for loss-free medium 
are shown by the following formula:

\begin{equation}
f_{mnp}=\left(c\sqrt{\frac{m^2}{a^2}+\frac{n^2}{b^2}+\frac{p^2}{l^2}}\right)/\left(2\sqrt{\varepsilon^{\prime}}\right),
\label{eq:f_mnp}
\end{equation}

{\noindent}where $m$, $n$, $p=0,1,2,\ldots$, $a$, $b$, $l$~--- the resonator 
dimensions along the Cartesian fields of coordinate axes ($x$, $y$, $z$), 
$c$~--- the velocity of electromagnetic waves in vacuum. 
For waves of type $H_{10}$ in a rectangular waveguide and for radiation 
propagation along axis $z$, resonators with the mode of oscillations 
$H_{10p}$ are used, where $p=1,2,3\ldots$. 
In the proposed technique, modes for $p>1$ should be used, i.e. $H_{102}$, 
$H_{103}$ etc. 
For this case, $m=1$, $n=0$ and $p=2,3\ldots$.

The respective resonance frequencies are:

\begin{equation}
f_{10p}=\left(c\sqrt{\frac{1}{a^2}+\frac{p^2}{l^2}}\right)/\left(2\sqrt{\varepsilon^{\prime}}\right).
\label{eq:f_10p}
\end{equation}

From Eq. (\ref{eq:f_10p}) for a certain mode of oscillations and based 
on the measurements of the resonance frequencies of an empty ($f_0$) 
and filled resonator, $\varepsilon_c^{\prime}$ of wetted medium is found:

\begin{equation}
\varepsilon_c^{\prime}=\left(\frac{f_0}{f_{10p}}\right)^2.
\label{eq:eps1_c}
\end{equation}

In case of losses to the medium, to the resonator walls and losses due 
to connection with the transmission waveguides, the value of the 
resonance frequency decreases, and $\varepsilon_c^{\prime}$ is found by 
the formula:

\begin{equation}
\varepsilon_c^{\prime}=\left(\frac{f_0}{f_{10p}}\right)^2\left(1-\frac{1}{4Q_c^2}\right),
\label{eq:eps1_cq}
\end{equation}

{\noindent}where $Q_c$~--- the quality factor for the resonator filled 
with the medium. 
For $Q_c\geq10$, which was observed in the experiment, the error of 
estimate $\varepsilon_c^{\prime}$ is less than $0.25$\%. 
Therefore, the value $\varepsilon_c^{\prime}$ for wetted medium filling 
the resonator was derived from formula (\ref{eq:eps1_c}).

To determine the dielectric parameters of water, a refraction model 
was used for a mixture of different dielectrics~\cite{BGHV1974}, according to 
which:

\begin{equation}
\sqrt{\dot{\varepsilon}_c}=\sqrt{\dot{\varepsilon}_m}V_m+(1-V_m)\sqrt{\dot{\varepsilon}_B},
\label{eq:eps_c}
\end{equation}

{\noindent}where $\dot{\varepsilon}_B$ is related to volume water in 
pores, $\dot{\varepsilon}_m$ is related to the structure consisting 
of solid matrix, empty space in pores and between the material grains, 
$V_m$~--- the relative volume of the medium, excluding liquid volume 
water. 

Equation (\ref{eq:eps_c}) contains complex variables and the unknown 
value $V_m$  and $\dot{\varepsilon}_m$. 
To solve equation (\ref{eq:eps_c}), additional information is introduced. 
It is assumed based on~\cite{LCh2012} that at temperature $-90\,^{\circ}$C free 
water freezes out. 
As its mass constitutes $\sim1$\% of the total mass of the substance, 
and the dielectric permittivity of the formed ice is close to value 
$\dot{\varepsilon}$ of the silicate material, we find $\dot{\varepsilon}_m$ 
from the resonance frequency measurements at a temperature lower than 
$-90\,^{\circ}$C within the error margin of one per cent. 
The real part $\dot{\varepsilon}_m$ is found from its equivalence 
$\left(f_0/f_{p max}\right)^2$, where $f_{p max}$~--- the resonance 
frequency lower than $-90\,^{\circ}$C. 
The imaginary part: 
$\varepsilon_m^{\prime\prime}=\varepsilon_m^{\prime}\left({\Delta}f_{p max}/f_{p max}\right)$. 
Here $\varepsilon_m^{\prime\prime}$~--- the effective loss factor, which 
takes into account losses to the dry medium and waveguide walls and 
losses due to connection with the waveguides, ${\Delta}f_{p max}$~--- 
resonance bandwidths at half power transmitting.

To find $V_m$, a priori information was found of the values of 
dielectric permittivity of water at $-18\,^{\circ}$C from~\cite{MWen2004}.  

From (\ref{eq:eps_c}) it follows: 

\begin{equation}
V_m=\frac{\sqrt{\dot{\varepsilon}_c}-\sqrt{\dot{\varepsilon}_B(-18)}}{\sqrt{\dot{\varepsilon}_m}-\sqrt{\dot{\varepsilon}_B(-18)}}.
\label{eq:V_m}
\end{equation}

$\dot{\varepsilon}_c$ is found from measurements at the temperature 
$-18\,^{\circ}$C. 
The real part $\varepsilon_c^{\prime}$ is determined from (\ref{eq:eps1_c}). 
$\varepsilon_c^{\prime\prime}$ is found from the measurements of the 
resonator's quality factor. 
Substituting the value $\varepsilon_c^{\prime\prime}$ for $-18\,^{\circ}$C 
in (\ref{eq:V_m}), we find $V_m$ and $(1-V_m)$.

The $Q_c$ is determined by the sum of two values: 

\begin{equation}
\frac{1}{Q_c}=\frac{1}{Q_a}+(tg\delta)_B,
\label{eq:Q_c}
\end{equation}

{\noindent}where $Q_c=f_{10p}/{\Delta}f_{10p}$, 
$(tg\delta)_B=\varepsilon_B^{\prime\prime}/\varepsilon_B^{\prime}$ refers 
to losses in volume water, $Q_a$ is the quality factor determined by 
losses to the resonator walls, connection to the waveguides and dry 
sorbent material. 
$Q_a=f_{p max}/{\Delta}f_{p max}$, determined from the measurements 
at a temperature lower than $-90\,^{\circ}$C. 
From (\ref{eq:Q_c}) we obtain:

\begin{equation}
\varepsilon_c^{\prime\prime}=\varepsilon_c^{\prime}\left(\frac{1}{Q_c}-\frac{1}{Q_a}\right).
\label{eq:eps2_c}
\end{equation}

Substituting the required values in (\ref{eq:eps_c}), we find $\dot{\varepsilon}_B$.

\section*{The procedure of measurements}

The microwave parameters of the resonator were measured using the 
Micran frequency characteristics analyzer. 
The scanning time for the entire frequency range from $6$ to $11$~GHz 
was $2$~s, and the number of the measured points in the frequency band 
was $5000$. 
The obtained values were processed on the computer, with dependences 
smoothed, the resonance curves approximated with bell-like functions 
and the resonance frequencies and width determined. 
The temperature of the medium investigated was measured with a 
thermocouple embedded in the opening of the resonator's wide wall. 

The temperature measurements were made using a climate chamber Espec 
SU-261, which allowed cooling of the resonator with the medium 
temperature down to $-60\,^{\circ}$C. 
To obtain lower temperatures and complete freezing of volume water, 
the chamber was switched off as the temperature reached $-60\,^{\circ}$C, 
and the resonator was cooled down with liquid nitrogen, followed by its 
slow heating. 
During those changes, hysteresis of the resonator characteristics 
($f_{10p}$ and ${\Delta}f_{10p}$) was observed. 
With the used cooling mode $\sim10\,^{\circ}$C/hr, nanoporous silicate 
sorbent materials with the pore diameter $3-4$~nm and their water 
content $4$\%, the observed hysteresis of the resonator characteristics 
in the range of temperatures $-30\ldots-50\,^{\circ}$C was 
$30\,^{\circ}$C. 
That allowed us to conclude that temperature of the supercooled water 
in the pores without water crystallization reached $-60\,^{\circ}$C. 
This conclusion was confirmed by the absence of the increase of 
$\varepsilon_B^{\prime\prime}$ at water cooling below $-23\,^{\circ}$C, 
at which ferroelectric ice~$0$ is formed. 
This crystal modification of ice results in the increase of the loss 
factor due to emergence of a highly conductive layer at the ice~--- 
dielectric boundary~\cite{W1963}. 

In the experiment, a transmission resonator was used, with the 
dimensions $a=23$~mm, $b=10$~mm, $l=43$~mm. 
Diaphragms with holes $5$~mm in diameter was placed in the plane $x$, $y$, 
where the cross section of the transmission  waveguides was 
$23\times10$~mm$^2$. 
The measured resonance frequencies of the empty resonators were found 
to be equal to: $f_{101}=7.420$~GHz, $f_{102}=9.805$~GHz, 
$f_{103}=12.23$~GHz.

\section*{Measurement results}

The above technique was applied to measurements in the range of 
frequencies $7.6\ldots9.7$~GHz at temperatures from $0\,^{\circ}$C 
to $-60\,^{\circ}$C. 
Their choice was connected with the experimental work~\cite{BCS1982}, where at the 
frequency $9.6$~GHz water was supercooled to $-18\,^{\circ}$C. 
The data from this study presented in~\cite{MWen2004} were used to update our 
measurements in determining the volume concentration of metastable 
water. 

In the experiment, KSKG silica gel made in Hong Kong with the average 
pore size $8$~nm and Acros silica gel manufactured in Belgium with 
the average pore size $6$ and $9$~nm were used. 
Measurements with a rectangular resonator were conducted simultaneously 
at two frequencies for the modes $H_{102}$ and $H_{103}$. 
For KSKG with the gravity water content $4$\%, resonances for the modes 
$H_{102}$ and $H_{103}$ were near the frequencies $7.6$~GHz and 
$9.7$~GHz. 

\begin{figure}[!t]
\centering
\includegraphics[width=0.9\columnwidth]{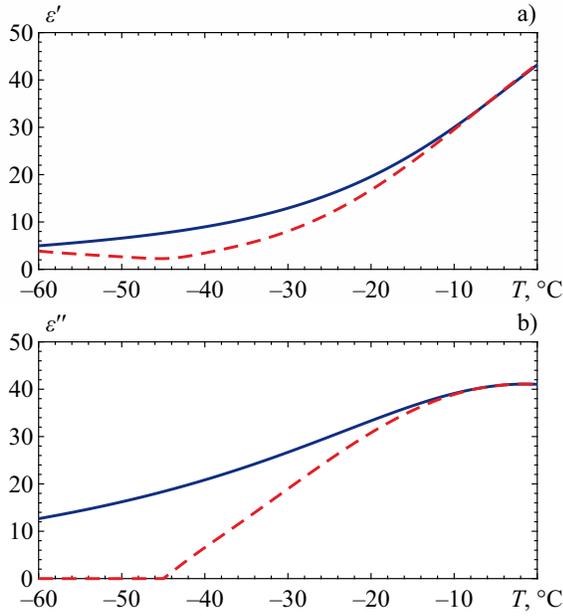}
\caption{Dependences: a) $\varepsilon^{\prime}(T)$ and b) $\varepsilon^{\prime\prime}(T)$ 
for volume water developed on the basis of measurements in a resonator 
near frequency $9.7$~GHz. The dashed curve~--- calculation results by 
formulae~\cite{MWen2004}.}
\label{fig:fig1}
\end{figure}

\begin{figure}[!t]
\centering
\includegraphics[width=0.9\columnwidth]{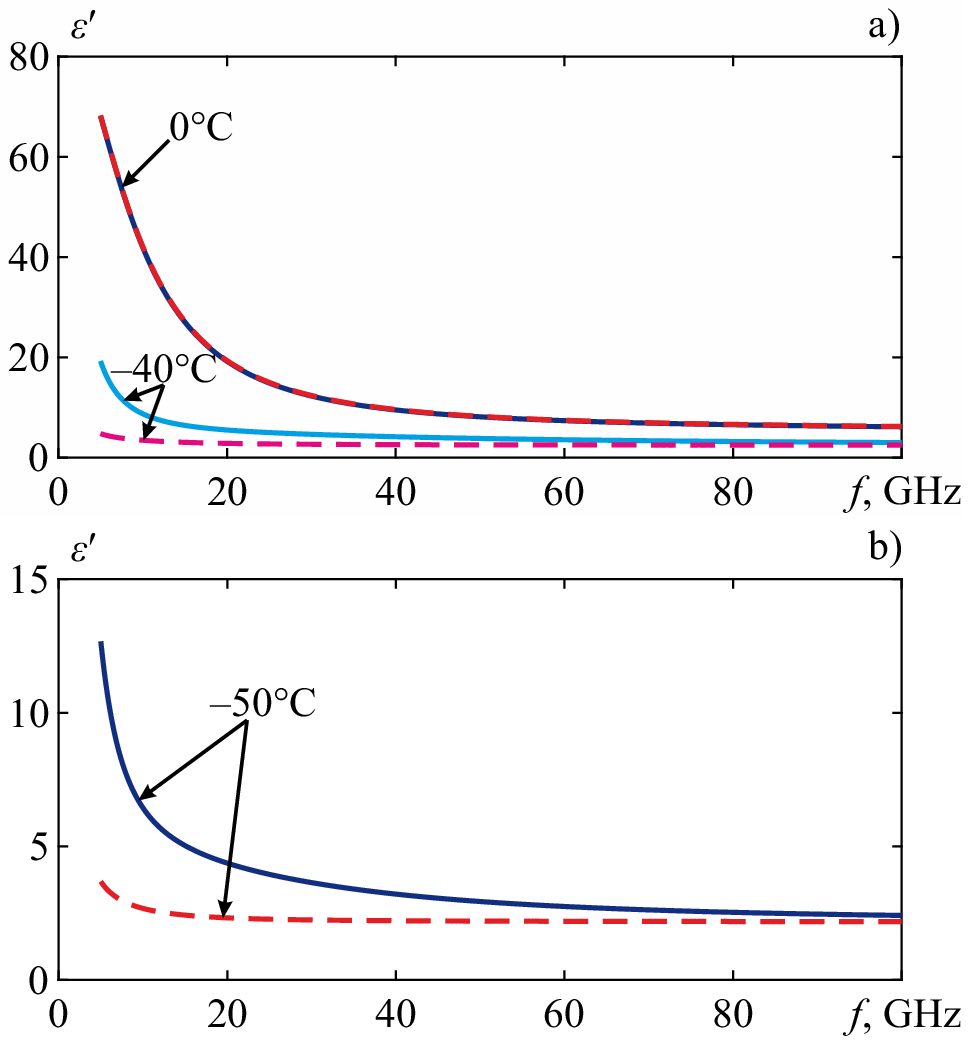}
\caption{Frequency dependences $\varepsilon^{\prime}$ of volume water 
at different temperatures. Dashed curve~--- calculations by formulae~\cite{MWen2004}. 
At $0\,^{\circ}$C two curves concur.}
\label{fig:fig2}
\end{figure}

\begin{figure}[!t]
\centering
\includegraphics[width=0.9\columnwidth]{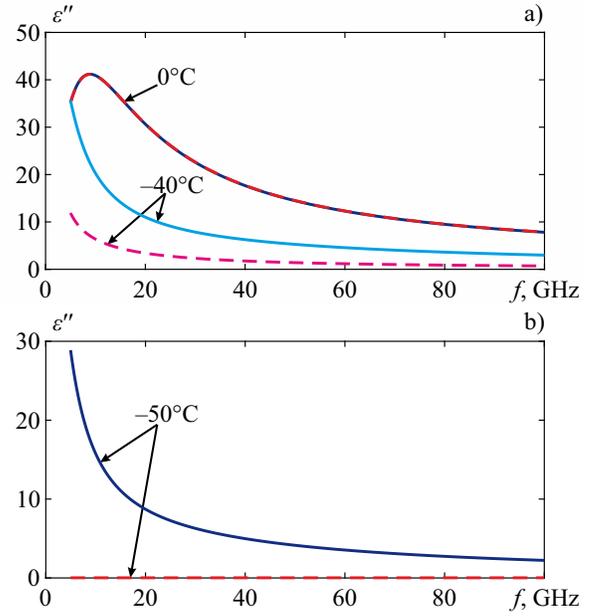}
\caption{Frequency dependencies $\varepsilon^{\prime\prime}$ of volume 
water at different temperatures. The dashed curve~--- calculations by 
formulae~\cite{MWen2004}. At $0\,^{\circ}$C two curves concur.}
\label{fig:fig3}
\end{figure}

\begin{figure}[!t]
\centering
\includegraphics[width=0.9\columnwidth]{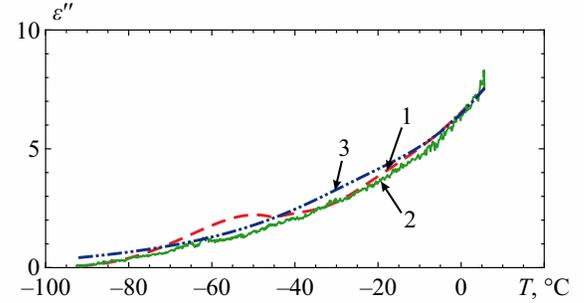}
\caption{Comparison of $\varepsilon^{\prime\prime}$ of volume water at 
the frequency of $125$~GHz: 1~--- dependence on the basis of the 
experiments in measuring the attenuation coefficient of wetted sorbent~\cite{BO2019b}
(dashed curve), 2~--- correction of the experimental results~\cite{BO2019b} 
using new data for $\varepsilon^{\prime}$ in this work, 3~--- calculations 
by formulae (\ref{eq:eps_T}, \ref{eq:MW}) with updated coefficients.}
\label{fig:fig4}
\end{figure}

The results of $\dot{\varepsilon}$ measurements after computer processing 
are shown in Figs.~\ref{fig:fig1}--\ref{fig:fig4}. 
As the data were processed, their approximation was performed with 
analytic dependences in accordance with~\cite{MWen2004}.

The formulae presented for the two-frequency Debye relaxation model 
look as follows for dielectric permittivity of pure water:

\begin{equation}
\dot{\varepsilon}\left(T\right)=\frac{\varepsilon_s\left(T\right)-\varepsilon_1\left(T\right)}{1+if/f_1\left(T\right)}+\frac{\varepsilon_1\left(T\right)-\varepsilon_\infty\left(T\right)}{1+if/f_2\left(T\right)}+\varepsilon_\infty\left(T\right),
\label{eq:eps_T}
\end{equation}

{\noindent}where $\varepsilon_1$ is the interim constant of dielectric 
permittivity, $f$ is frequency, $\varepsilon_s$~--- static dielectric 
constant, $f_1$ and $f_2$~--- the first and second relaxation frequencies. 
For $\varepsilon_s$, $\varepsilon_1$, $\varepsilon_\infty$, $f_1$, $f_2$, 
there are the following formulae~\cite{MWen2004}: 

\begin{eqnarray}
\label{eq:MW}
	\varepsilon_s\left(T\right)=\frac{3.70886\cdot10^4-8.2168\cdot10^1T}{4.21854\cdot10^2+T}\\
	\varepsilon_1\left(T\right)=a_0+a_1T+a_2T^2\nonumber\\
	f_1\left(T\right)=\frac{A+T}{a_3+a_4T+a_5T^2}\nonumber\\
	\varepsilon_\infty\left(T\right)=a_6+a_7T\nonumber\\
	f_2\left(T\right)=\frac{A+T}{a_8+a_9T+a_{10}T^2},\nonumber
\end{eqnarray}

{\noindent}where $f$ is expressed in GHz, $T$~--- in degrees Celsius.

For the purpose of their use in the region of negative temperatures, 
dependences of two relaxation frequencies $f_{1,2}\left(T\right)$ were 
specified. 
In~\cite{MWen2004}, assuming the impact of the Widom line of water for the 
temperature of $-45\,^{\circ}$C~\cite{GAAA2016, W1963, HBAS2012} parameter $A$ was chosen 
to be equal to $45$. 
As $T$ was expressed in degrees Celsius, at that relaxation temperatures 
turned into zero. 
In our measurements, we observed an electromagnetic response at lower 
temperatures, too.

Based on the assumption that in cold water its vitrification occurs at 
the temperature of approximately $-130\,^{\circ}$C~\cite{HBAS2012}, parameter $A$ 
was selected to be equal to $130$. 
After adjustment of the coefficients in formulae (\ref{eq:MW}) to 
achieve compliance of $\varepsilon^{\prime}$  and $\varepsilon^{\prime\prime}$ 
with the experimental values at the temperatures below $0\,^{\circ}$C, 
the following coefficient values were found: the table I contains coefficients 
in formulae (\ref{eq:MW}).

\begin{table}
\caption{Coefficients in formulae (\ref{eq:MW})}
\label{tab:tab_1}
\centering
\begin{tabular}{|c|c|c|c|}
\hline
\bfseries $i$ & \bfseries $a_i$ & \bfseries $i$ & \bfseries $a_i$\\
\hline
	0 &	5.6500E00 &	6 &	3.6143E00\\
\hline	
	1 &	1.6960E-02 &	7 &	2.8841E-02\\
\hline	
	2 &	-1.4810E-04 &	8 &	3.9208E-01\\
\hline	
	3 &	1.4627E01 &	9 &	-3.2094E-03\\
\hline	
	4 &	-4.2926E-01 &	10 &	7.6578E-04\\
\hline	
	5 &	7.5714E-03 & & \\ 
\hline
\end{tabular}
\end{table}

Shown in Fig. \ref{fig:fig1} are dependences $\varepsilon^{\prime}$ and 
$\varepsilon^{\prime\prime}$ on temperature for frequency $9.7$~GHz. 
To compare, the results of computations using formulae~\cite{MWen2004} are provided. 
At the temperature below $-45\,^{\circ}$C, in~\cite{MWen2004} the values 
$\varepsilon^{\prime\prime}$ are not determined, therefore, they were 
taken to be equal to zero.
  
Frequency dependences $\varepsilon^{\prime}$ and $\varepsilon^{\prime\prime}$ 
for different temperatures are shown in Figs. \ref{fig:fig2}, 
\ref{fig:fig3}. 
  
The obtained values of $\varepsilon^{\prime}$ were used for adjusting 
the results for $\varepsilon^{\prime\prime}$ at the frequencies of the 
millimeter band by the measurements of the attenuation coefficient 
provided in~\cite{BO2019b}. Fig.~\ref{fig:fig4} demonstrates the temperature 
dependences $\varepsilon^{\prime\prime}$ from~\cite{BO2019b} for the frequency 
of $125$~GHz. 
 
\section*{Discussion of results}

The main difficulty in conducting measurements of $\dot{\varepsilon}$ 
consists in eliminating the distortions of resonance curves in the form 
of oscillations on the curve slopes. 
The distortions are related to inhomogeneities emerging in wetted micro 
disperse medium. 
In wetted medium with the water content above $4$\%, it is practically 
impossible to make measurements. 
This feature made computer calculations more difficult due to the 
difficulty of determining the precise values of the resonance frequency 
and of the width of the resonance curve. 
In smoothing of the measured resonance curves: 1~--- by averaging of 
curves by neighboring points in the band of frequencies; 2~--- by 
Gaussian curves; 3~--- by the Lorentz dependence, variations of the 
final results were observed ($\varepsilon^{\prime}$) $\sim\pm25$\% 
from the mean value. 
The smoothest dependences $\varepsilon^{\prime}(T)$ were obtained at 
averaging in the bandwidths $70$~MHz (at $100$~points), followed by 
approximation by the Gaussian function. 

To obtain dependences $\dot{\varepsilon}$ on frequency, formulae 
from~\cite{MWen2004} were specified for the two-frequency Debye model for pure 
volume water. 
The main idea was in correct selection of the temperature dependence 
of relaxation frequencies on temperature. 
The application of the two-frequency model has certain reasons due to 
development of a two-structure model of water, which considers the 
fluctuating clusters of water of high-density (HDL) and of low-density 
(LDL)~\cite{GAAA2016}. 
These structures are in complex interaction. 
Their concentration is determined by temperature, and at cooling below 
$-50\,^{\circ}$C LDL begins to prevail. 
At temperature around $-130\,^{\circ}$C the relaxation time becomes 
equal to $100$~s (vitrification begins), hence, in Debye model near 
this temperature we can consider the relaxation frequencies to turn 
into zero. 

Validation of formulae (\ref{eq:eps_T}, \ref{eq:MW}) for 
the available measurement data of $\varepsilon^{\prime\prime}$ 
at millimeter frequencies~\cite{BO2019a, BO2019b} demonstrates good 
compliance of the experimental results obtained by using two different 
methods. 
A number of factors influence the accuracy of the results, apart from 
the measurement error. 
It is known that water density decreases as it is cooled from $+4\,^{\circ}$C 
to $-70\,^{\circ}$C~\cite{Mallamace18387}. 
This effect is compensated to a certain degree by selection of 
coefficients in the temperature dependence of the relaxation times. 
A more complicated issue consists in compliance of the parameters of pore 
water and of volume water (for macro volumes), as well as the error 
introduced by strongly bound water. 
For strongly bound water, the relaxation frequencies are in the range 
of $0.1$~GHz, but this water may influence the value of $\varepsilon^{\prime}$ 
at the frequencies of the gigahertz order. 

In certain cases, if crystallization begins at temperatures below 
$-23\,^{\circ}$C, not ice Ih or Ic but ice~$0$ may be formed in wetted 
medium~\cite{QAS2014, BO2017}. 
The peculiar feature of this ice is that it is ferroelectric and has 
strong influence on the microwave characteristics of the medium~\cite{BO2017}. 
Ice~$0$ may be formed on the grain surface or in the space in-between 
the sorbent grains. 
It is likely that it caused the failure of the experiments in wetted 
media with the water content higher than $3-4$\%. 

The accuracy of determining the values $\dot{\varepsilon}$ by formulae 
(\ref{eq:MW}) for volume water at temperatures below $-20\,^{\circ}$C 
is approximately equal to $30$\%. 
In order to raise the degree of this accuracy, measurements are 
required to be made in a broader range of frequencies and the special 
relaxation and absorption mechanisms have to be investigated. 
For example, in~\cite{BO2019a, BO2019b} the increase of $\varepsilon^{\prime\prime}$ 
near $-45\,^{\circ}$C was investigated due to the impact of the specific 
behavior of water on the Widom line~\cite{HBAS2012}. 
The Widom line is related to the existence of the second critical point 
of water. 
At it, fluctuations of water density and entropy drastically rise, and 
a number of the thermodynamic characteristics of water change 
(specific heat capacity at constant pressure, coefficients of volume 
expansion and of isothermal compressibility).

It is to be noted that the obtained results proved to be applicable, 
based on the comparison results, to the temperature $20\,^{\circ}$C and 
the frequency $200$~GHz.

\section*{Conclusions}

1.	A method is proposed for measuring relative dielectric 
permittivity of supercooled water in the microwave range for 
temperatures from $-20\,^{\circ}$C to $-60\,^{\circ}$C. 
It is based on the special properties of water contained in 
nanoporous silicate materials, in which water is close to 
metastable volume water for its parameters. 
In this technique, in order to eliminate the impact of 
inhomogeneities resulting from emergence of clusters at 
migration of water due to temperature gradients and phase 
transitions, measurements are conducted in resonators which 
have relatively large volumes and a low degree of wetting of 
the materials. 
In this case, the impact of the inhomogeneities is averaged, and 
distortions in the measured electromagnetic characteristics of the 
media under study become eliminated. 
The values $\varepsilon^{\prime}$ and $\varepsilon^{\prime\prime}$ 
near the frequencies $7.6$~GHz and $9.6$~GHz were measured, and 
analytical dependence $\dot{\varepsilon}(T)$ in the range of 
$-20\ldots-60\,^{\circ}$C was obtained. 
The obtained dependence $\dot{\varepsilon}(T)$ may be spread to the range 
of frequencies $7\ldots200$~GHz and temperatures $-90\ldots+20\,^{\circ}$C, 
which the comparison results showed.

2.	The technique proposed may ensure approximately $30$\% accuracy 
of measuring dielectric permittivity of supercooled water. 
However, a question remains regarding correspondence of the parameters 
of water contained in the pores of a material to those of the ideal 
volume water. 
The accuracy of determining the dielectric characteristics of water 
depends on a number of parameters: the proportion of bound water and 
the breakdown of its characteristics by  the water layers adjacent to 
the surfaces of the medium boundaries; the specific details of the 
sample cooling techniques; the physical-chemical characteristics of 
the sorbent matrix; the values of electric conductivity of the layers 
at the medium interfaces and certain other parameters. 
Therefore, the expected accuracy of determining the values of 
dielectric permittivity in relation to ideal volume water may 
significantly vary in any method of measurement. 
This relates also to the previously conducted measurements using 
microemulsions of water~\cite{BCS1982}. 
In our experiments involving two different types of silica gels, the 
difference between the values of $\varepsilon^{\prime}$ reached $25$\%.

3.	The measurements performed refer to cold water in the region of 
``no man's land''. 
This region is interesting due to the most conspicuous anomaly of water, 
decrease of its density when temperature changes from $+4\,^{\circ}$C 
to $-70\,^{\circ}$C~\cite{Mallamace18387}. 
In its turn, this anomaly is related to the structural features of 
liquid water: interaction and mutual transformations of clusters LDL 
and HDL, and the existence of the second critical point. 
The specific features of the dielectric permittivity of supercooled 
water in porous media may be used in microwave spectroscopy of the 
processes of water freezing (melting) in closed space. 
For example, it has been found that at freezing of water below 
$-23\,^{\circ}$C, ferroelectric ice~$0$ is formed. 
Its appearance results in the rise of electromagnetic losses, which 
may be used for the study of the chemical processes in cooling of 
natural disperse media due to the change of the chemical potential 
of the inclusions. 
Equally interesting is the issue of investigating the microwave 
characteristics of water near the Widom line using the proposed 
technique (at the temperatures $-45\ldots-53\,^{\circ}$C and the 
pressure $0.1\ldots100$~MPa). 
At present, there are little experimental results on this issue.

\bibliographystyle{apsrev4-1}
\bibliography{LGC_Bib}

\begin{thebibliography}{21}%
\makeatletter
\providecommand \@ifxundefined [1]{%
 \@ifx{#1\undefined}
}%
\providecommand \@ifnum [1]{%
 \ifnum #1\expandafter \@firstoftwo
 \else \expandafter \@secondoftwo
 \fi
}%
\providecommand \@ifx [1]{%
 \ifx #1\expandafter \@firstoftwo
 \else \expandafter \@secondoftwo
 \fi
}%
\providecommand \natexlab [1]{#1}%
\providecommand \enquote  [1]{``#1''}%
\providecommand \bibnamefont  [1]{#1}%
\providecommand \bibfnamefont [1]{#1}%
\providecommand \citenamefont [1]{#1}%
\providecommand \href@noop [0]{\@secondoftwo}%
\providecommand \href [0]{\begingroup \@sanitize@url \@href}%
\providecommand \@href[1]{\@@startlink{#1}\@@href}%
\providecommand \@@href[1]{\endgroup#1\@@endlink}%
\providecommand \@sanitize@url [0]{\catcode `\\12\catcode `\$12\catcode
  `\&12\catcode `\#12\catcode `\^12\catcode `\_12\catcode `\%12\relax}%
\providecommand \@@startlink[1]{}%
\providecommand \@@endlink[0]{}%
\providecommand \url  [0]{\begingroup\@sanitize@url \@url }%
\providecommand \@url [1]{\endgroup\@href {#1}{\urlprefix }}%
\providecommand \urlprefix  [0]{URL }%
\providecommand \Eprint [0]{\href }%
\providecommand \doibase [0]{http://dx.doi.org/}%
\providecommand \selectlanguage [0]{\@gobble}%
\providecommand \bibinfo  [0]{\@secondoftwo}%
\providecommand \bibfield  [0]{\@secondoftwo}%
\providecommand \translation [1]{[#1]}%
\providecommand \BibitemOpen [0]{}%
\providecommand \bibitemStop [0]{}%
\providecommand \bibitemNoStop [0]{.\EOS\space}%
\providecommand \EOS [0]{\spacefactor3000\relax}%
\providecommand \BibitemShut  [1]{\csname bibitem#1\endcsname}%
\let\auto@bib@innerbib\@empty
\bibitem [{\citenamefont {Meissner}\ and\ \citenamefont
  {Wentz}(2004)}]{MWen2004}%
  \BibitemOpen
  \bibfield  {author} {\bibinfo {author} {\bibfnamefont {T.}~\bibnamefont
  {Meissner}}\ and\ \bibinfo {author} {\bibfnamefont {F.~J.}\ \bibnamefont
  {Wentz}},\ }\href@noop {} {\bibfield  {journal} {\bibinfo  {journal} {IEEE
  Trans. Geosci. Remote Sens.}\ }\textbf {\bibinfo {volume} {42}},\ \bibinfo
  {pages} {1836} (\bibinfo {year} {2004})}\BibitemShut {NoStop}%
\bibitem [{\citenamefont {Rosenkranz}(2015)}]{Ros2015}%
  \BibitemOpen
  \bibfield  {author} {\bibinfo {author} {\bibfnamefont {P.~W.}\ \bibnamefont
  {Rosenkranz}},\ }\href@noop {} {\bibfield  {journal} {\bibinfo  {journal}
  {IEEE Trans. Geosci. Remote Sens.}\ }\textbf {\bibinfo {volume} {53}},\
  \bibinfo {pages} {1387} (\bibinfo {year} {2015})}\BibitemShut {NoStop}%
\bibitem [{\citenamefont {Turner}\ \emph {et~al.}(2016)\citenamefont {Turner},
  \citenamefont {Kneifel},\ and\ \citenamefont {Cadeddu}}]{TKC2016}%
  \BibitemOpen
  \bibfield  {author} {\bibinfo {author} {\bibfnamefont {D.~D.}\ \bibnamefont
  {Turner}}, \bibinfo {author} {\bibfnamefont {S.}~\bibnamefont {Kneifel}}, \
  and\ \bibinfo {author} {\bibfnamefont {M.~P.}\ \bibnamefont {Cadeddu}},\
  }\href@noop {} {\bibfield  {journal} {\bibinfo  {journal} {J. Atmos. Oceanic
  Technol.}\ }\textbf {\bibinfo {volume} {33}},\ \bibinfo {pages} {33}
  (\bibinfo {year} {2016})}\BibitemShut {NoStop}%
\bibitem [{\citenamefont {Bertolini}\ \emph {et~al.}(1982)\citenamefont
  {Bertolini}, \citenamefont {Cassettari},\ and\ \citenamefont
  {Salvetti}}]{BCS1982}%
  \BibitemOpen
  \bibfield  {author} {\bibinfo {author} {\bibfnamefont {D.}~\bibnamefont
  {Bertolini}}, \bibinfo {author} {\bibfnamefont {M.}~\bibnamefont
  {Cassettari}}, \ and\ \bibinfo {author} {\bibfnamefont {G.}~\bibnamefont
  {Salvetti}},\ }\href@noop {} {\bibfield  {journal} {\bibinfo  {journal} {J.
  Chem. Phys.}\ }\textbf {\bibinfo {volume} {76}},\ \bibinfo {pages} {3285}
  (\bibinfo {year} {1982})}\BibitemShut {NoStop}%
\bibitem [{\citenamefont {Bordonskii}\ \emph {et~al.}(2019)\citenamefont
  {Bordonskii}, \citenamefont {Orlov},\ and\ \citenamefont {Krylov}}]{BOK2019}%
  \BibitemOpen
  \bibfield  {author} {\bibinfo {author} {\bibfnamefont {G.~S.}\ \bibnamefont
  {Bordonskii}}, \bibinfo {author} {\bibfnamefont {A.~O.}\ \bibnamefont
  {Orlov}}, \ and\ \bibinfo {author} {\bibfnamefont {S.~D.}\ \bibnamefont
  {Krylov}},\ }\href@noop {} {\bibfield  {journal} {\bibinfo  {journal}
  {Journal of Communications Technology and Electronics}\ }\textbf {\bibinfo
  {volume} {64}},\ \bibinfo {pages} {375} (\bibinfo {year} {2019})}\BibitemShut
  {NoStop}%
\bibitem [{\citenamefont {Bordonskiy}\ and\ \citenamefont
  {Orlov}(2019{\natexlab{a}})}]{BO2019a}%
  \BibitemOpen
  \bibfield  {author} {\bibinfo {author} {\bibfnamefont {G.~S.}\ \bibnamefont
  {Bordonskiy}}\ and\ \bibinfo {author} {\bibfnamefont {A.~O.}\ \bibnamefont
  {Orlov}},\ }\href@noop {} {\bibfield  {journal} {\bibinfo  {journal}
  {Preprint arXiv: 1901.03979 [cond-mat.soft]}\ ,\ \bibinfo {pages} {8}}
  (\bibinfo {year} {2019}{\natexlab{a}})}\BibitemShut {NoStop}%
\bibitem [{\citenamefont {Limmer}\ and\ \citenamefont
  {Chandler}(2012)}]{LCh2012}%
  \BibitemOpen
  \bibfield  {author} {\bibinfo {author} {\bibfnamefont {D.~T.}\ \bibnamefont
  {Limmer}}\ and\ \bibinfo {author} {\bibfnamefont {D.}~\bibnamefont
  {Chandler}},\ }\href@noop {} {\bibfield  {journal} {\bibinfo  {journal} {J.
  Chem. Phys.}\ }\textbf {\bibinfo {volume} {137}},\ \bibinfo {pages}
  {044509/1} (\bibinfo {year} {2012})}\BibitemShut {NoStop}%
\bibitem [{\citenamefont {Bordonskiy}\ and\ \citenamefont
  {Orlov}(2019{\natexlab{b}})}]{BO2019b}%
  \BibitemOpen
  \bibfield  {author} {\bibinfo {author} {\bibfnamefont {G.~S.}\ \bibnamefont
  {Bordonskiy}}\ and\ \bibinfo {author} {\bibfnamefont {A.~O.}\ \bibnamefont
  {Orlov}},\ }\href@noop {} {\bibfield  {journal} {\bibinfo  {journal}
  {Izvestiya, Atmospheric and Oceanic Physics}\ }\textbf {\bibinfo {volume}
  {55}},\ \bibinfo {pages} {1005} (\bibinfo {year}
  {2019}{\natexlab{b}})}\BibitemShut {NoStop}%
\bibitem [{\citenamefont {Ellison}(2007)}]{Ell2007}%
  \BibitemOpen
  \bibfield  {author} {\bibinfo {author} {\bibfnamefont {W.~J.}\ \bibnamefont
  {Ellison}},\ }\href@noop {} {\bibfield  {journal} {\bibinfo  {journal} {J.
  Chem. Phys. Ref. Data}\ }\textbf {\bibinfo {volume} {36}},\ \bibinfo {pages}
  {1} (\bibinfo {year} {2007})}\BibitemShut {NoStop}%
\bibitem [{\citenamefont {Menshikov}\ \emph {et~al.}(2017)\citenamefont
  {Menshikov}, \citenamefont {Menshikov},\ and\ \citenamefont
  {Fedichev}}]{MMF2017}%
  \BibitemOpen
  \bibfield  {author} {\bibinfo {author} {\bibfnamefont {L.~I.}\ \bibnamefont
  {Menshikov}}, \bibinfo {author} {\bibfnamefont {P.~L.}\ \bibnamefont
  {Menshikov}}, \ and\ \bibinfo {author} {\bibfnamefont {P.~O.}\ \bibnamefont
  {Fedichev}},\ }\href@noop {} {\bibfield  {journal} {\bibinfo  {journal}
  {Journal of Experimental and Theoretical Physics}\ }\textbf {\bibinfo
  {volume} {125}},\ \bibinfo {pages} {1173} (\bibinfo {year}
  {2017})}\BibitemShut {NoStop}%
\bibitem [{\citenamefont {Castrillon}\ \emph {et~al.}(2011)\citenamefont
  {Castrillon}, \citenamefont {Giovambattista}, \citenamefont {Arsay},\ and\
  \citenamefont {Debenedetti}}]{CGAD2011}%
  \BibitemOpen
  \bibfield  {author} {\bibinfo {author} {\bibfnamefont {S.~R.-V.}\
  \bibnamefont {Castrillon}}, \bibinfo {author} {\bibfnamefont
  {N.}~\bibnamefont {Giovambattista}}, \bibinfo {author} {\bibfnamefont
  {I.~A.}\ \bibnamefont {Arsay}}, \ and\ \bibinfo {author} {\bibfnamefont
  {P.~G.}\ \bibnamefont {Debenedetti}},\ }\href@noop {} {\bibfield  {journal}
  {\bibinfo  {journal} {J. Phys. Chem. C}\ }\textbf {\bibinfo {volume} {115}},\
  \bibinfo {pages} {4624} (\bibinfo {year} {2011})}\BibitemShut {NoStop}%
\bibitem [{\citenamefont {Solveyra}\ \emph {et~al.}(2011)\citenamefont
  {Solveyra}, \citenamefont {Llave}, \citenamefont {Scherlis},\ and\
  \citenamefont {Molinero}}]{SLSM2011}%
  \BibitemOpen
  \bibfield  {author} {\bibinfo {author} {\bibfnamefont {E.~G.}\ \bibnamefont
  {Solveyra}}, \bibinfo {author} {\bibfnamefont {E.}~\bibnamefont {Llave}},
  \bibinfo {author} {\bibfnamefont {D.~A.}\ \bibnamefont {Scherlis}}, \ and\
  \bibinfo {author} {\bibfnamefont {V.}~\bibnamefont {Molinero}},\ }\href@noop
  {} {\bibfield  {journal} {\bibinfo  {journal} {J. Phys. Chem. B}\ }\textbf
  {\bibinfo {volume} {115}},\ \bibinfo {pages} {14196} (\bibinfo {year}
  {2011})}\BibitemShut {NoStop}%
\bibitem [{\citenamefont {Gallo}\ \emph {et~al.}(2016)\citenamefont {Gallo},
  \citenamefont {Amann-Winkel}, \citenamefont {Angell}, \citenamefont
  {Anisimov}, \citenamefont {Caupin}, \citenamefont {Chakravarty},
  \citenamefont {Lascaris}, \citenamefont {Loerting}, \citenamefont
  {Panagiotopoulos}, \citenamefont {Russo}, \citenamefont {Sellberg},
  \citenamefont {Stanley}, \citenamefont {Tanaka}, \citenamefont {Vega},
  \citenamefont {Xu},\ and\ \citenamefont {Pettersson}}]{GAAA2016}%
  \BibitemOpen
  \bibfield  {author} {\bibinfo {author} {\bibfnamefont {P.}~\bibnamefont
  {Gallo}}, \bibinfo {author} {\bibfnamefont {K.}~\bibnamefont {Amann-Winkel}},
  \bibinfo {author} {\bibfnamefont {C.~A.}\ \bibnamefont {Angell}}, \bibinfo
  {author} {\bibfnamefont {M.~A.}\ \bibnamefont {Anisimov}}, \bibinfo {author}
  {\bibfnamefont {F.}~\bibnamefont {Caupin}}, \bibinfo {author} {\bibfnamefont
  {C.}~\bibnamefont {Chakravarty}}, \bibinfo {author} {\bibfnamefont
  {E.}~\bibnamefont {Lascaris}}, \bibinfo {author} {\bibfnamefont
  {T.}~\bibnamefont {Loerting}}, \bibinfo {author} {\bibfnamefont {A.~Z.}\
  \bibnamefont {Panagiotopoulos}}, \bibinfo {author} {\bibfnamefont
  {J.}~\bibnamefont {Russo}}, \bibinfo {author} {\bibfnamefont {J.~A.}\
  \bibnamefont {Sellberg}}, \bibinfo {author} {\bibfnamefont {H.~E.}\
  \bibnamefont {Stanley}}, \bibinfo {author} {\bibfnamefont {H.}~\bibnamefont
  {Tanaka}}, \bibinfo {author} {\bibfnamefont {C.}~\bibnamefont {Vega}},
  \bibinfo {author} {\bibfnamefont {L.}~\bibnamefont {Xu}}, \ and\ \bibinfo
  {author} {\bibfnamefont {L.~G.~M.}\ \bibnamefont {Pettersson}},\ }\href@noop
  {} {\bibfield  {journal} {\bibinfo  {journal} {Chem. Rev.}\ }\textbf
  {\bibinfo {volume} {116}},\ \bibinfo {pages} {7463} (\bibinfo {year}
  {2016})}\BibitemShut {NoStop}%
\bibitem [{\citenamefont {Webber}(2010)}]{W2010}%
  \BibitemOpen
  \bibfield  {author} {\bibinfo {author} {\bibfnamefont {J.~B.~W.}\
  \bibnamefont {Webber}},\ }\href@noop {} {\bibfield  {journal} {\bibinfo
  {journal} {Progress in Nuclear Magnetic Resonance Spectroscopy}\ }\textbf
  {\bibinfo {volume} {56}},\ \bibinfo {pages} {78} (\bibinfo {year}
  {2010})}\BibitemShut {NoStop}%
\bibitem [{\citenamefont {M$\ddot{\mathrm{a}}$tzler}\ and\ \citenamefont
  {Wegmuller}(1987)}]{MatWeg1987}%
  \BibitemOpen
  \bibfield  {author} {\bibinfo {author} {\bibfnamefont {C.}~\bibnamefont
  {M$\ddot{\mathrm{a}}$tzler}}\ and\ \bibinfo {author} {\bibfnamefont
  {U.}~\bibnamefont {Wegmuller}},\ }\href@noop {} {\bibfield  {journal}
  {\bibinfo  {journal} {J. Phys. D: Appl. Phys.}\ }\textbf {\bibinfo {volume}
  {20}},\ \bibinfo {pages} {1623} (\bibinfo {year} {1987})}\BibitemShut
  {NoStop}%
\bibitem [{\citenamefont {Birchak}\ \emph {et~al.}(1974)\citenamefont
  {Birchak}, \citenamefont {Gardner}, \citenamefont {Hipp},\ and\ \citenamefont
  {Victor}}]{BGHV1974}%
  \BibitemOpen
  \bibfield  {author} {\bibinfo {author} {\bibfnamefont {J.~R.}\ \bibnamefont
  {Birchak}}, \bibinfo {author} {\bibfnamefont {L.~G.}\ \bibnamefont
  {Gardner}}, \bibinfo {author} {\bibfnamefont {J.~W.}\ \bibnamefont {Hipp}}, \
  and\ \bibinfo {author} {\bibfnamefont {J.~M.}\ \bibnamefont {Victor}},\
  }\href@noop {} {\bibfield  {journal} {\bibinfo  {journal} {Proc. IEEE}\
  }\textbf {\bibinfo {volume} {62}},\ \bibinfo {pages} {93} (\bibinfo {year}
  {1974})}\BibitemShut {NoStop}%
\bibitem [{\citenamefont {Widom}(1963)}]{W1963}%
  \BibitemOpen
  \bibfield  {author} {\bibinfo {author} {\bibfnamefont {B.}~\bibnamefont
  {Widom}},\ }\href@noop {} {\bibfield  {journal} {\bibinfo  {journal} {J.
  Chem. Phys.}\ }\textbf {\bibinfo {volume} {39}},\ \bibinfo {pages} {2808}
  (\bibinfo {year} {1963})}\BibitemShut {NoStop}%
\bibitem [{\citenamefont {Holten}\ \emph {et~al.}(2012)\citenamefont {Holten},
  \citenamefont {Bertrand}, \citenamefont {Anisimov},\ and\ \citenamefont
  {Sengers}}]{HBAS2012}%
  \BibitemOpen
  \bibfield  {author} {\bibinfo {author} {\bibfnamefont {V.}~\bibnamefont
  {Holten}}, \bibinfo {author} {\bibfnamefont {C.~E.}\ \bibnamefont
  {Bertrand}}, \bibinfo {author} {\bibfnamefont {M.~A.}\ \bibnamefont
  {Anisimov}}, \ and\ \bibinfo {author} {\bibfnamefont {J.~V.}\ \bibnamefont
  {Sengers}},\ }\href@noop {} {\bibfield  {journal} {\bibinfo  {journal} {J.
  Chem. Phys.}\ }\textbf {\bibinfo {volume} {136}},\ \bibinfo {pages} {094507}
  (\bibinfo {year} {2012})}\BibitemShut {NoStop}%
\bibitem [{\citenamefont {Mallamace}\ \emph {et~al.}(2007)\citenamefont
  {Mallamace}, \citenamefont {Branca}, \citenamefont {Broccio}, \citenamefont
  {Corsaro}, \citenamefont {Mou},\ and\ \citenamefont {Chen}}]{Mallamace18387}%
  \BibitemOpen
  \bibfield  {author} {\bibinfo {author} {\bibfnamefont {F.}~\bibnamefont
  {Mallamace}}, \bibinfo {author} {\bibfnamefont {C.}~\bibnamefont {Branca}},
  \bibinfo {author} {\bibfnamefont {M.}~\bibnamefont {Broccio}}, \bibinfo
  {author} {\bibfnamefont {C.}~\bibnamefont {Corsaro}}, \bibinfo {author}
  {\bibfnamefont {C.-Y.}\ \bibnamefont {Mou}}, \ and\ \bibinfo {author}
  {\bibfnamefont {S.-H.}\ \bibnamefont {Chen}},\ }\href@noop {} {\bibfield
  {journal} {\bibinfo  {journal} {Proceedings of the National Academy of
  Sciences}\ }\textbf {\bibinfo {volume} {104}},\ \bibinfo {pages} {18387}
  (\bibinfo {year} {2007})}\BibitemShut {NoStop}%
\bibitem [{\citenamefont {Quigley}\ \emph {et~al.}(2014)\citenamefont
  {Quigley}, \citenamefont {Alf$\grave{\mathrm{e}}$},\ and\ \citenamefont
  {Slater}}]{QAS2014}%
  \BibitemOpen
  \bibfield  {author} {\bibinfo {author} {\bibfnamefont {D.}~\bibnamefont
  {Quigley}}, \bibinfo {author} {\bibfnamefont {D.}~\bibnamefont
  {Alf$\grave{\mathrm{e}}$}}, \ and\ \bibinfo {author} {\bibfnamefont
  {B.}~\bibnamefont {Slater}},\ }\href@noop {} {\bibfield  {journal} {\bibinfo
  {journal} {J. Chem. Phys.}\ }\textbf {\bibinfo {volume} {141}},\ \bibinfo
  {pages} {161102/1} (\bibinfo {year} {2014})}\BibitemShut {NoStop}%
\bibitem [{\citenamefont {Bordonskiy}\ and\ \citenamefont
  {Orlov}(2017)}]{BO2017}%
  \BibitemOpen
  \bibfield  {author} {\bibinfo {author} {\bibfnamefont {G.~S.}\ \bibnamefont
  {Bordonskiy}}\ and\ \bibinfo {author} {\bibfnamefont {A.~O.}\ \bibnamefont
  {Orlov}},\ }\href@noop {} {\bibfield  {journal} {\bibinfo  {journal} {JETP
  Letters}\ }\textbf {\bibinfo {volume} {105}},\ \bibinfo {pages} {492}
  (\bibinfo {year} {2017})}\BibitemShut {NoStop}%
\end{thebibliography}%

\end{document}